\documentclass[11pt,a4paper]{article}
\usepackage{jcappub}
\usepackage{bm}
\usepackage{dcolumn}
\usepackage{graphicx}

\newif\ifAMStwofonts
\AMStwofontstrue

\def\wpdot{W_{{\dot \phi}}}
\def\wph{W_{ \phi}}
\def\wv{W_{ \rm V}}

\def\zcos{z_{_{\rm cos}}}

\def\vnabla{\bm{\nabla}}

\def\gsim{~\rlap{$>$}{\lower 1.0ex\hbox{$\sim$}}}

\def\simpropto{\lower.2ex\hbox{$\; \buildrel \propto \over \sim \;$}}
\def\ltsim{\lower.5ex\hbox{$\; \buildrel < \over \sim \;$}}
\def\gtsim{\lower.5ex\hbox{$\; \buildrel > \over \sim \;$}}
\def\ltsim{\lower.5ex\hbox{$\; \buildrel < \over \sim \;$}}
\def\gtsim{\lower.5ex\hbox{$\; \buildrel > \over \sim \;$}}

\def\vnabla{{\bf \nabla}}

\def\sv{\Theta{^{\rm V}}}
\def\sw{\Theta{^{\rm \Phi}}}
\def\sisw{\Theta{^{\rm { \dot \Phi}}}}

\def\tsw{{\tilde \Theta}{^{\rm \Phi}}}
\def\tsisw{{\tilde \Theta}{^{\rm { \dot \Phi}}}}

\def\fsk{f_{_{\rm sky}}}

\def\kms{\mbox{km\,s$^{-1}$}}

\def\dd{\,{\rm d}}

\def\kms{\ {\rm km\,s^{-1}}}
\def\hmpc{\ {\rm h^{-1}Mpc}}

\def\dd{{\rm d}}

\def\ln{{\rm ln}}

\def\pmb#1{\setbox0=\hbox{#1}%
\kern-.025em\copy0\kern-\wd0
\kern.05em\copy0\kern-\wd0
\kern-.025em\raise.0433em\box0}

\def\vr{\pmb{$r$}}
\def\vn{\pmb{$n$}}
\def\hvn{\hat {\vr}}

\def\vk{\pmb{$k$}}

\def\simlt{\lower.5ex\hbox{$\; \buildrel < \over \sim \;$}}
\def\simgt{\lower.5ex\hbox{$\; \buildrel > \over \sim \;$}}

\def\vnabla{\pmb{$\nabla$}}

\newcommand{\beq}{\begin{equation}}
\newcommand{\eeq}{\end{equation}}
\def\beqa{\begin{eqnarray}}
\def\eeqa{\end{eqnarray}}
\def\fixit#1{}
\def\hmpc{h^{-1}\,{\rm Mpc}}

\def\dd{{\rm d}}

\notoc

\begin{document}

\title{A direct probe of cosmological power spectra of the peculiar velocity field and the gravitational lensing magnification from photometric redshift surveys}

\author[a,b]{Adi Nusser,}
\author[c,d,e]{Enzo Branchini,}
\author[a]{and Martin Feix}

\affiliation[a]{Department of Physics, Israel Institute of Technology - Technion, Haifa 32000, Israel}
\affiliation[b]{Asher Space Science Institute, Israel Institute of Technology - Technion, Haifa 32000, Israel}
\affiliation[c]{Department of Physics, Universit\`a Roma Tre, Via della Vasca Navale 84, Rome 00146, Italy}
\affiliation[d]{INFN Sezione di Roma 3, Via della Vasca Navale 84, Rome 00146, Italy}
\affiliation[e]{INAF, Osservatorio Astronomico di Brera, Via Brera 28, Milano 20121, Italy}

\emailAdd{adi@physics.technion.ac.il}
\emailAdd{branchin@fis.uniroma3.it}
\emailAdd{mfeix@physics.technion.ac.il}
    
\abstract{The cosmological peculiar velocity field (deviations from the pure Hubble flow) of matter carries significant information on dark energy,
dark matter and the underlying theory of gravity on large scales. Peculiar motions of galaxies introduce systematic deviations between
the observed galaxy redshifts $z$ and the corresponding cosmological redshifts $z_{_{\rm cos}}$. A novel method for estimating the angular
power spectrum of the peculiar velocity field based on observations of galaxy redshifts and apparent magnitudes $m$ (or equivalently fluxes)
is presented. This method exploits the fact that a mean relation between $z_{_{\rm cos}}$ and $m$ of galaxies can be derived from all galaxies
in a redshift-magnitude survey. Given a galaxy magnitude, it is shown that the $z_{_{\rm cos}}(m)$ relation yields its cosmological redshift
with a $1\sigma $ error of $\sigma_z\sim 0.3$ for a survey like Euclid ($\sim 10^9$ galaxies at $z\lesssim 2$), and can be used to constrain
the angular power spectrum of $z-z_{_{\rm cos}}(m)$ with a high signal-to-noise ratio. At large angular separations corresponding to $l\lesssim 15$,
we obtain significant constraints on the power spectrum of the peculiar velocity field. At $15 \lesssim l\lesssim 60$, magnitude shifts in the
$z_{_{\rm cos}}(m)$ relation caused by gravitational lensing magnification dominate, allowing us to probe the line-of-sight integral of the
gravitational potential. Effects related
to the environmental dependence in the luminosity function can easily be computed and their contamination removed from the estimated power spectra.
The amplitude of the combined velocity and lensing power spectra at $z\sim 1$ can be measured with $\lesssim 5\%$ accuracy.}

\arxivnumber{1207.5800}

\maketitle
\section{Introduction}
\label{sec:int}
Dark matter, dark energy, and the theory of gravitation dictate the evolution of large-scale structure in the Universe. The physical conditions
allowing for the formation of galaxies, ultimately lead to a bias between the distribution of galaxies and the underlying mass density. However,
large-scale motions of galaxies are most certainly locked to the peculiar velocity field associated with the gravitational tug of the total underlying
mass fluctuations. This assumes that gravity is the only relevant large-scale force and neglects the contribution from decaying linear modes. Despite
the bias between the distribution of galaxies and the underlying matter field, the clustering properties of galaxies have been the main tool for
testing cosmological models. Currently planned galaxy surveys will allow us to quantify the clustering of galaxies on hundreds of comoving Mpcs and to
even measure coherent distortions of galaxy images which arise from gravitational lensing by the foreground matter.

On the other hand, the peculiar motions of galaxies have traditionally been less successful as a cosmological tool, and there are several reasons
for that. Neglecting other potentially important effects (see section \ref{method} for details), peculiar velocities are approximately equal to the redshifts less the corresponding
Hubble expansion recession velocities. The latter require direct measurements of galaxy distances which are available for only a small fraction
of galaxies. Presently, the number of galaxies with measured distances is several orders of magnitude below that of galaxies in redshift surveys used
for clustering studies. Furthermore, although the underlying peculiar velocity is an honest tracer of the general matter flow, the inference of
peculiar velocities from observational data is plagued with observational biases \cite{lyn88}. Traditional peculiar velocity catalogs are expected
to improve within the next few years, but it remains questionable how well observational biases will be controlled, especially at large distances.
An alternative probe of the peculiar velocity field may be astrometric measurements of galaxies by the Gaia space mission \cite{perryman01,nbdgaia}.
This probe is essentially free of the classic biases contaminating traditional peculiar velocity measurements, but it is also limited to nearby galaxies
within $\sim 100\hmpc$.

Here we describe a method for deriving strong constraints on the power spectrum of the galaxies' peculiar velocity field independent of
conventional direct distance measurements, which are prone to systematic errors, and any biasing relation between galaxies and mass. The
method is an extension of the approaches we have recently proposed \citep{NBDL,ND11a,BDN12}, and it relies on using the observed fluxes of galaxies as a proxy for their cosmological distance \cite{TYS1979}. Although
this most basic distance indicator is very noisy, the large number of galaxies available in future surveys will allow one to beat down this noise
to a sufficiently low level. Planned galaxy redshift surveys such as Euclid \cite{euclidL,EuclidRB} will probe the structure of the Universe over
thousands of (comoving) Mpcs, comprising $\lesssim 10^8$--$10^9$ galaxies at $z\sim 1$ and beyond. These observations will provide redshifts and fluxes,
$z$ and $f$, respectively.
The observed redshifts deviate from the cosmological redshifts which would be observed in a purely homogeneous universe. The
large number of galaxies and the large sky coverage of these surveys can be used to derive a mean global relation between the mean redshift of a
galaxy and its apparent magnitude $m=-2.5 \log f + {\rm const}$. We interpret this relation as yielding the cosmological redshift $z_{_{\rm cos}}(m)$
for a given apparent magnitude $m$. Angular power spectra of the difference $z_i- z_{_{\rm cos}}(m_i)$ between the observed redshift $z_i$ of a
galaxy and its expected cosmological redshift $z_{_{\rm cos}}$ should contain valuable information, mainly on the peculiar velocity field which
is the main cosmological source for $z_i- z_{_{\rm cos}}(m_i)$. In this work, we will show that the velocity power spectrum on large scales of a
few 100 Mpcs could be constrained with significant signal-to-noise ratio ($S/N$) at the effective depth of the survey. Another contribution to
$z_i- z_{_{\rm cos}}(m_i)$ results from the time evolution of the gravitational potential along the photon path, but is significantly smaller than
that induced by peculiar velocities as we will show below.
There are two additional, indirect effects which modify the relation $z_{_{\rm cos}}(m)$ along a given line of sight. The first effect is related
to the environmental dependence between galaxy luminosities and the large-scale structure in which they reside. Since this dependence is closely
connected to the underlying density field, it can be self-consistently removed in our analysis. The second effect is caused by gravitational lensing
magnification which changes the apparent magnitudes of galaxies in a given direction. This latter contribution could actually be very rewarding since
gravitational lensing provides a direct probe of the underlying mass distribution. Considering the analysis presented below, we will therefore treat
it as part of the sought signal. 

The paper is structured as follows: We begin with a detailed description of the method and its application to galaxy redshift surveys in section
\ref{method}. In section \ref{sec:tpk}, we consider predictions for the standard $\Lambda$CDM model and discuss the method's viability as well as
its expected performance. Finally, we present our conclusions in section \ref{sec:cnl}. For clarity, some of the technical material is given separately
in an appendix. In the following, we adopt the standard notation. The matter density and the cosmological constant in units of the critical density
are denoted by $\Omega$ and $\Lambda$, respectively. The scale factor $a$ is normalized to unity at the present time ($t=t_0$), and the Hubble
function is defined as $H=\dot{a}/a$. Further, $r=c \int_{t}^{t_0} dt'/a(t')$ will be the comoving distance to an object and $z$ its corresponding
redshift, assuming a homogeneous and isotropic cosmological background. Throughout the paper, the subscript ``$0$'' will refer to quantities given
at $t=t_0$, and a dot symbol denotes partial derivatives with respect to time $t$, i.e. $\dot{A}\equiv\partial A/\partial t$.

\section{Methodology}
\label{method}
In an inhomogeneous universe, the observed redshift $z$ of a given object differs from its cosmological redshift $z_{_{\rm cos}}$ (defined for the
unperturbed background). The relative difference between these redshifts, $\Theta \equiv (z-z_{_{\rm cos}})/(1+z_{_{\rm cos}})$, can be expressed as
\cite{SW}
\begin{equation}
\label{eq:Theta}
\Theta=\frac{V(t,r)}{c} - \frac{\Phi(t,r)}{c^2} - \frac{2}{c^2}\int_{t(r)}^{t_0}dt \frac{\partial \Phi\left\lbrack\hvn r(t),t\right\rbrack}{\partial t},
\end{equation}
where $\hvn $ is a unit vector along the line-of-sight to the object. Here the radial peculiar velocity $V$ and the usual metric potential $\Phi$ are
assumed as relative to their present-day values at $r=0$ ($t=t_{0}$). The first and second terms on the right-hand side of eq. \eqref{eq:Theta} are
the Doppler and gravitational shifts, respectively, while the third term describes the energy change of light as it passes through a time-varying
gravitational potential. Note that this third term is equivalent to the late-time integrated Sachs-Wolfe effect experienced by photons of the cosmic
microwave background (CMB). In what follows, we will denote the three terms as $\sv$, $ \sw$ and $ \sisw$, respectively. Also, we will consider
angular power spectra (equivalent to angular correlations) of $\Theta$ on large scales where the corresponding signal is significant only relative to
the expected error. Throughout this paper, we therefore rely on linear theory in a $\Lambda\rm CDM$ model where perturbations on all scales grow at
the same rate. Introducing $D(t)$ as the growth rate of the underlying mass density contrast $\delta=\rho/\bar \rho-1$, linear theory yields the
well-known relations 
\begin{align}
\delta(t,\vr) &= D(t) \delta_0(\vr),\\
\Phi(\vr,t) &= \frac{D(t)}{a}\Phi_0(\vr),\\
\label{eq:linv} V(t,\vr) &= -\frac{2}{3}\frac{a \dot D(t)}{\Omega_0 H_0^2}\frac{\partial \Phi_0}{\partial r}, 
\end{align}
where $D(t_0)=1$. The second relation is obtained from the first using Poisson's equation, i.e. $\vnabla_r^2\Phi = 3 H_0^2\Omega_0\delta/2a $. For the
special case $\Omega_0=1$, we have $D=a$ and fluctuations in the gravitational potential remain constant with time.

\subsection{Cosmological redshift versus apparent magnitude}
\label{methodsub1}
We aim to derive the field $\Theta$ sampled at the positions of all galaxies in a flux limited redshift survey covering a significant region of
the sky together with a large number of galaxies. As an example, we consider the planned Euclid redshift survey \cite{euclidL,EuclidRB}. In order
to obtain an estimate of $\Theta_{i}$ for each galaxy, we shall use apparent galaxy magnitudes (or equivalently fluxes) as a proxy to the cosmological
redshift or distance. Although this most trivial distance indicator is very noisy, we will see that the high number of available galaxies allows
one to beat down its scatter.

The mean cosmological redshift $\zcos(m)$ corresponding to a given apparent magnitude $m$ is
\begin{equation}
\zcos(m)=\bar z_{_{\rm cos}}=\int_{z_1}^{z_2}\zcos  P(\zcos|m) \dd \zcos ,
\end{equation}
and the root mean square (rms) scatter around this relation is 
\begin{equation}
\sigma_z(m)=\int_{z_1}^{z_2} (\zcos -\bar z_{_{\rm cos}} )^2P(\zcos|m)\dd \zcos ,
\end{equation}
where $z_1$ and $z_2$ are the limiting redshifts in the survey, and $P(\zcos|m)$ denotes the probability that a galaxy with measured apparent
magnitude $m$ has a cosmological redshift $\zcos$. The $\zcos(m)$ relation can be linked to certain characteristics of the galaxy survey. The
luminosity of a galaxy at $\zcos$ is $L=4\pi f d_L^2$ where $d_L(\zcos)$ is the luminosity distance to the galaxy, and its absolute magnitude
is defined as $M=-2.5 \log L +{\rm const}=m-5 \log d_L$. Let $\Phi(M)$ be the underlying luminosity function such that $\Phi \dd M $ is the
number density of galaxies within the magnitude interval $[M,M+\dd M]$. Generally, the function $\Phi(M)$ depends on cosmic time, but in favor
of a simplified description, we will assume for the moment that it varies little throughout the depth of the survey considered. Note that it
is trivial to include a time-dependent evolutionary term in $\Phi(M)$. The probability $P(\zcos|m)$ satisfies
\begin{equation}
\label{eq:pcos}
P(\zcos|m)\propto P(m|\zcos)n(\zcos),
\end{equation}
where $n(\zcos) $ is the underlying mean number density of galaxies at $\zcos$. In the absence of galaxy population evolution, we have
\begin{equation}
n(\zcos) \propto d_c(\zcos)^2 \frac{\dd d_c(\zcos)}{\dd \zcos},
\end{equation}
where $d_c$ is the comoving distance from the observer to $\zcos$. The relation in eq. \eqref{eq:pcos} can be easily derived using Bayes'
theorem which yields
\begin{equation} 
P(\zcos |m)P(m) = P(m|\zcos )P(\zcos),
\end{equation}
where $P(m)$ can be directly estimated from observations, $P(\zcos)$ is proportional to $n(\zcos)$, and $P(m|\zcos)$ is related to $\Phi(M)$ through  
\begin{equation}
P(M|\zcos) = \frac{\Phi(M)}{\int_{-\infty}^{M_l(\zcos)}\Phi(M)\dd M}.
\end{equation}
Here $M_l = m_l-5\log d_L(\zcos)$ is the absolute magnitude which corresponds to the limiting apparent magnitude $m_l$ of the galaxy survey.

Note that the scatter of $\zcos$ about the mean relation is not Gaussian. However, the associated $1\sigma$ error in the derived angular
power spectra depends only on the rms quantity $\sigma_z$ (see section \ref{methodsub2} for a detailed discussion of systematic errors).
Moreover, the central limit theorem implies that the errors in the power spectra tend to be Gaussian. Considering the actual observations,
the two quantities $\zcos(m)$ and $\sigma_z(m)$ can be estimated from the full survey by dividing the data in magnitude bins without
actually computing $P(\zcos|m)$. In this case, $\sigma_z(m)$ will include a positive contribution from the cosmological deviations, but
this is overwhelmed by both the intrinsic scatter in the $\zcos(m)$ relation and the uncertainties in the photometric redshifts.
The underlying assumption is that the sought cosmological deviations between observed and cosmological redshifts cancel out when all
galaxies of the entire survey are used. Clearly, a global constant mode should persist in this procedure, but we shall neglect it in this
paper, assuming that it does not affect modes in the power spectra on smaller scales. Once we obtain $\zcos(m)$, we compute
\begin{equation}
\Theta_{i} = \frac{z_{i} - \zcos(m_{i})}{1 + z_{i}}
\end{equation}
for all galaxies in the survey. Note that in the denominator of the above, we have used $z_{i}$ instead of $\zcos(m_{i})$. This
substitution is consistent at linear order and motivated by the fact that $z_{i}$ is actually a better estimate of the true $\zcos $
than $\zcos(m_{i})$ which additionally includes deviations from the actual value of $\zcos$ due to large random errors with an rms of
$\sigma_z$. 

Environmental dependences of the luminosity distribution on the large-scale density field may systematically shift all $\Theta_i$ for galaxies
lying in the direction of a certain line of sight. However, we will show below that the resulting signal contamination is small and in any case,
it can be removed from the correlations since information on the underlying density field at the relevant scales will be directly available from
the observations (see section \ref{sec:env}).

\subsection{Expectations for Euclid}
\label{methodsub2}
As a test case for a future survey, we consider Euclid which aims at measuring photometric and spectroscopic redshifts of galaxies with 
$\zcos \sim 1$ over $15,000$ $\rm deg^2$. More details on the Euclid mission can be found in \cite{EuclidRB}. Here we focus on the
photometric redshift survey since we aim at good statistics rather than precise redshift measurements. Photometric redshifts will be
measured with an error of $\sigma_{\rm phot}\le 0.05(1+z_{\rm phot})$ for $\sim  30$ galaxies per arcmin$^2$, with $\zcos \ge 0.7$ and
magnitudes in the broad R+I+Z band (550--920 nm) RIZ$_{\rm AB} \le 24.5$. Estimates of the photometric redshifts will rely on three
near-infrared (NIR) bands (Y, J, and H in the range 0.92--2.0 $\mu$m) for objects with Y$_{\rm AB}\le 24$, J$_{\rm AB}\le 24$, and
H$_{\rm AB}\le 24$. These will further be complemented by ground-based photometry in the visible bands derived from public data or
through engaged collaborations.

The expected number density of Euclid galaxies with measured photometric redshift can be parametrized as follows \citep{theoryWG}: 
\begin{equation}
\label{eq:nz}
n(z_{\rm phot})\propto z_{\rm phot}^2 {\rm e}^{-(z_{\rm phot}/z_0)^{3/2}},
 \end{equation}
where $z_{0} = z_{\rm mean}/1.412$ is the peak of the redshift distribution and $z_{\rm mean}$ the median. Here we assume that
$z_{\rm mean} =0.9$. We have compared the expected distribution $n(z)$ of eq. \eqref{eq:nz} to that measured from galaxies with 
H$_{\rm AB}\le 24$ and RIZ$_{\rm AB} \le 24.5$ in the zCOSMOS catalog \cite{Ilbert09,Mck10,Bielby11}. Indeed, we have found that eq.
\eqref{eq:nz} does provide a good fit to the data in the range $0.7 < z_{\rm phot} < 2.0$ which we will consider in our analysis.

\begin{figure} 
\centering
\includegraphics[scale=0.7]{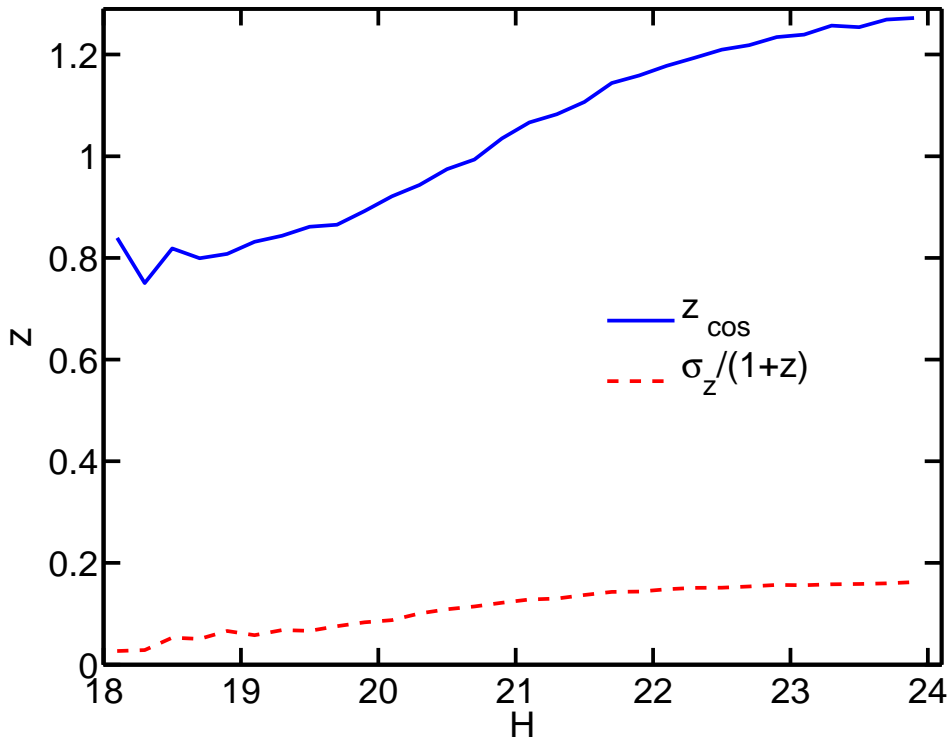}
\caption{The mean relation $\zcos(m)$ and the corresponding rms scatter $\sigma_z(m)/(1+z)$ for Euclid galaxies with
RIZ$_{\rm AB} \le 24.5$, H$_{\rm AB}\le 24$, and photometric redshifts in the range $0.7 < z_{\rm phot} < 2.0$: The shown results
are based on zCOSMOS data.}
\label{fig:one}
\end{figure}

The observed $z_{\rm phot}$-H relation of zCOSMOS galaxies has been used to derive $\bar z_{\rm phot}(\rm H)$ and $\sigma_{z_{\rm phot}}(\rm H)$
in different H-magnitude bins. To match Euclid constraints, we have only considered galaxies with RIZ$_{\rm AB} \le 24.5$,
H$_{\rm AB}\le 24$, and $0.7 < z_{\rm phot} < 2.0$. The additional constraints Y$_{\rm AB}\le 24$ and J$_{\rm AB}\le 24$ do not
significantly modify our results and have therefore not been enforced. Our results for the zCOSMOS data are shown in figure \ref{fig:one}.
The (blue) solid curve and the (red) dashed curve represent the expected $\zcos(H)$ and $\sigma_z(H)$ for Euclid galaxies, respectively.
As can be directly read off the figure, the expected scatter in $\zcos(H)$ is $\sigma_z(H)\lesssim 0.3$.
Errors on the measured redshift, i.e. $\sigma_{\rm phot}$, dominate the scatter for H$<20$. They will be added in quadrature to
determine the effective scatter in the $\zcos$-H relation.

Since our aim is to estimate the angular correlation properties of $\Theta_{i}$, we are concerned with all potential sources of systematic
errors that are coherent over large angular scales. Random errors in the H-band photometry and $z_{\rm phot}$ can induce systematic errors
as a result of the galaxies' non-uniform distribution in the $z_{\rm phot}$-H plane. However, the resulting offsets bear no angular
coherence and can be safely ignored in our analysis. The lack of angular correlations also characterizes systematic errors induced by a
gross misestimate of $z_{\rm phot}$ which are commonly known as ``catastrophic errors".

Angular-dependent systematic errors may arise when calibrating the photometry across a large area of the sky. The angular structure of
these errors generally depends on the survey strategy and, considering ground-based observations, its interplay with the atmospheric
conditions at the telescope's site. In current surveys such as SDSS, the relative photometric errors are already on the order of 1\%
(10 mmag) and have only little angular structure \cite{Padma2008}. Clearly, these will be further reduced in next-generation surveys, especially in those that will be based in space. Systematic errors of this kind may propagate into zeropoint offsets affecting the
estimate of $z_{\rm phot}$. To assess the significance of this effect for our analysis, we have performed a number of simulations in
which we have computed $z_{\rm phot}$ after introducing photometric offsets of about $1$\% in different bands, using various templates
of spectral energy distributions for different galaxy types. As a reference case, we have considered a set of $1000$ galaxies at $z=1$
observed with the Euclid filters RIZ, Y, J, and H. A small, but sizable zeropoint offset $\Delta z_{\rm phot} = 0.01$ can be obtained
if the photometric offset runs smoothly from $-5$ mmag in RIZ to $5$ mmag in the H band. It is rather unlikely that such a
configuration will ever occur, but even so, there are plenty of reasons to ignore these systematic errors. First of all, the amplitude
and probability of $\Delta z_{\rm phot}$ decreases dramatically with the number of used filters. For instance, if additional photometry
in the visible bands $g$, $r$, $i$, and $z$ is considered (which is expected to be the case for Euclid), then $\Delta z_{\rm phot}$ drops
by a factor of approximately $10$. Second, we have found that the offset's amplitude can be further reduced by selecting homogeneous
subsamples of objects. Third, $\Delta z_{\rm phot}$ turns out proportional to the photometry offset, and will decrease when sub-percent
calibration accuracy is achieved. Finally, $z_{\rm phot}$ can be independently calibrated in different areas of the sky with the help of
spectroscopic redshift information, which may be used to reveal a possible angular correlation among errors.

\subsection{Spherical harmonics decomposition for discrete noisy data} 
\label{methodsub3}
Much of the analysis presented below is very similar to previous work done in the lensing community, and before that, in the context of the
CMB \cite{Knox1995, Hobson1996}. As usual, angular power spectra are defined in terms of the spherical harmonics $Y_{lm}(\hvn)$. For an
all-sky continuous field $f(\hvn)$, the decomposition is
\begin{equation}
f_{lm} = \int d\Omega f(\hvn)Y_{lm}(\hvn),\quad f({\hvn}) = \sum_{l=0}^\infty\sum_{m=-l}^{+l} f_{lm}Y^*_{lm}(\hvn),
\end{equation}
In our case, however, we have to deal with partial sky coverage (around $30\%$ of the sky for Euclid) and consider that the field is sampled
at discrete points given by the galaxy positions. The limited sky coverage could formally be described by an appropriate masking of the sphere
\cite{Peeb73,Peeb80}. However, the description in terms of masks will unnecessarily complicate the notation and somewhat obscure the physical
interpretation of the results. We shall therefore resort to a simplified description and assume that we are provided with a survey covering
$4\pi\fsk $ steradians of the sky. For each degree $l$, we assume that there are $(2l+1)\fsk$ independent modes, instead of $2l+1$ for full
sky coverage. Considering modes with angular resolution much smaller than the extent of the survey, we then have
\begin{equation}
\label{eq:rules}
\begin{split}
&\int d\Omega Y_{lm}(\hvn) Y^*_{l^{\prime}m^{\prime}}(\hvn) = \fsk\delta^K_{ll^{\prime}}\delta^K_{mm^{\prime}},\\
&\sum_m Y_{lm}(\hvn) Y^*_{lm}(\hvn^{\prime}) = \frac{(2l+1)\fsk}{4\pi}P_l(\hvn\cdot \hvn^{\prime}),
\end{split}
\end{equation}
where $P_l$ is the Legendre polynomial of degree $l$. Throughout the paper, the angular integration is carried out only over the observed part
of the sky, and the number of terms in the sum of the second relation is $(2l+1)\fsk$. As is obvious, these relations should be understood to
hold in the approximate sense.\footnote{We emphasize that a future analysis of the real data should properly account for the lack of a full sky
coverage.}

Considering a function $f(\hvn)$ with limited sky coverage in the continuous limit, we define
\begin{equation}
\label{eq:flm}
f_{lm} = \frac{1}{ \fsk^{1/2}} \int d \Omega  f(\hvn) Y_{lm}(\hvn),
\end{equation}
where the integration is again taken over the observed region only. The angular power spectrum $C_{l}$ is defined as the variance of the
$f_{lm}$'s and given by
\begin{equation}
C_{l} = \langle\left\vert f_{lm} \right\vert^{2}\rangle_{_{\rm ens}},
\end{equation}
where the average is taken over many different realizations of a field with the same power spectrum as $f$. For brevity, we will use the symbol
$\langle\cdot\rangle$ without any subscript to refer to this kind of averaging. Up to cosmic variance, the values of $C_{l}$ obtained from the ensemble average
$\langle\left\vert f_{lm} \right\vert^{2}\rangle$ should equal those computed by averaging over the $(2l+1)\fsk$ modes for each $l$. Thus, the angular power
spectrum may be estimated as
\begin{equation}
C_l= \langle\left\vert f_{lm} \right\vert^{2}\rangle_{_{m}}\equiv \frac{1}{(2l+1)\fsk}\sum_m \left\vert f_{lm} \right\vert^2 . 
\end{equation}
In the following, we shall interchange between these different kinds of averaging whenever appropriate. The factor $1/\fsk^{1/2}$ in eq.
\eqref{eq:flm} and the rules in eq. \eqref{eq:rules} guarantee that 
\begin{equation}
\label{eq:CLC}
C(\cos\theta) = \sum\limits_l C_l \frac{2l+1}{4\pi} P_l(\cos\theta),\quad C_{l} = \int d\Omega C(\cos\theta) P_l(\cos\theta),
\end{equation}
which can be shown by decomposing $P_{l}$ into $Y_{lm} $ according to the second relation in eq. \eqref{eq:rules} and assuming that
$C(\theta) $ is negligible for angular separations larger than the extent of the survey (see appendix \ref{sec:proof}).

For a discrete sampling of $f$ at the positions of $N$ galaxies distributed over the observed part of the sky, we write \cite{afsh04}
\begin{equation}
\sum_i^N f(\hvn_i) \approx \int d\Omega n(\hvn)f(\hvn) \approx \bar n \int d\Omega f(\hvn),
\end{equation}
where $n(\hvn)$ is the projected number density of objects and $\bar n=N/(\fsk 4\pi)$ is the corresponding mean number density over
the observed part of the sky. Here we have assumed that $f$ itself depends on the density contrast $\delta=n/\bar n-1$, so that the 
last step applies to linear order in the fluctuations. Again, the angular integration is carried out only over the observed part of
the sky. In analogy to eq. \eqref{eq:flm}, we further define
\begin{equation}
\label{eq:fdef}
f_{lm} = \frac{1}{\bar n \fsk^{1/2}} \sum_{i=1}^N f(\hvn_i) Y_{lm}(\hvn_i) .
\end{equation}
In this case, one can show that the angular power spectrum takes the form (see appendix \ref{sec:shotn}) 
\begin{equation}
C_{l} = \langle\left\vert f_{lm} \right\vert^{2}\rangle - \frac{\sigma_f^2}{\bar n},
\end{equation}
where the second term represents the contribution of shot noise due to the discrete sampling of $f$ and $\sigma_f^2=\sum_i f(\hvn_i)^2/N$.
To estimate the values of $C_{l}$, we therefore use
\begin{equation}
\label{eq:CLf}
C_{l} = \langle\left| f_{lm}\right|^{2}\rangle_{_{m}} - \frac{\sigma_f^2}{\bar n}.
\end{equation} 
If $f_{i} = S(\hvn_i) + \epsilon_{i}$ where $S$ is an underlying cosmological signal and $\epsilon_{i}$ is an uncorrelated random
error, this simply reflects the fact that $C_{l} = \langle\left| S_{lm} \right|^{2}\rangle$. The expected error in this estimate of $C_l$ is given
by (again, see appendix \ref{sec:shotn})
\begin{equation}
\label{eq:SIGMA}
\Sigma^{2} = \frac{2}{(2l+1)\fsk}\left(\frac{\sigma_f^2}{\bar n}+C_{l}\right)^{2}
\end{equation}
which includes contributions from both shot noise $\Sigma_{\rm sn}$ and cosmic variance $\Sigma_{\rm cv}$,
\begin{equation}
\Sigma_{\rm sn} = \sqrt{\frac{2}{(2l+1)\fsk}}\frac{\sigma_f^2}{\bar{n}},\quad \Sigma_{\rm cv} = \sqrt{\frac{2}{(2l+1)\fsk}}C_l
\end{equation}
The variance of the scatter $\sigma^2_i$,
\begin{equation}
\label{eq:sigi_def}
\sigma_{i}^{2} = \langle\epsilon_{i}^{2}\rangle = \frac{\sigma_{\rm phot}^{2}(z_i) + \sigma_{z}^{2}(m_i)}{(1+z_{i})^{2}},
\end{equation}
depends on both magnitude and redshift. Note that the factor of $(1+z_{i})$ arises from the definition of $\Theta$ in eq. \eqref{eq:Theta}.
Considering the application to real data, it is prudent to weight each galaxy according to the $\sigma_{i}$ in the sum of eq. \eqref{eq:fdef}.
To minimize the effects of shot noise, we weight each galaxy by a factor of $w_{i}$ which is given by
\begin{equation}
\label{eq:w}
w_{i}^{2} = \frac{N \sigma_{i}^{-2}}{\sum_{j} \sigma_{j}^{-2}}.
\end{equation}
This particular weighting scheme yields
\begin{equation}
\label{eq:sigf}
\sigma_{f}^{2} = \frac{N}{\sum_{j}\sigma_j^{-2}},
\end{equation}
where we have assumed that that the underlying signal makes a negligible contribution to $\sigma_{f}^{2}$. The weighting does not affect the
ensemble average of $\lvert f_{lm}\rvert^{2}$, and its net effect is that $\sigma_{f}$ should be computed as given by eq. \eqref{eq:sigf},
i.e. it merely reduces shot noise errors (for additional details, see appendix \ref{sec:shotn}). In principle, one may use any weighting scheme.

\subsection{Environmental dependences in the luminosity functions and magnification by gravitational lensing}
\label{sec:env}
So far, we have assumed that the systematic shifts in the $\zcos(m)$ relation for galaxies in a given direction are solely due to the terms appearing
on the right-hand side of eq. \eqref{eq:Theta}. However, additional shifts may arise from changes in the mean magnitudes due to large-scale density
fluctuations in a given direction, i.e. environmental dependences and evolution in the luminosity function, and the magnification effect caused by
gravitational lensing. In what follows, we shall denote their contribution as $\Theta^{\rm env}$ and $\Theta^{\rm lens}$, respectively. Both effects
result in a magnitude shift which translates into additional correlated deviations of the estimated $\zcos$ from its mean relation. To model the
contribution of these effects in the correlation function of $\Theta$, we need to translate a magnitude shift $\Delta m$ into a corresponding shift
$\Delta z$. This can directly be read off the (blue) solid curve $\zcos(m)$ shown in figure \ref{fig:one} in which a shift $\Delta m=0.2$ leads to mean
shift $\Delta\zcos\approx 0.017$.\footnote{Note that although more complicated schemes are possible, we adopt this mean relation for simplicity.}

To quantify the impact of environmental dependences, we assume that the systematic shift in galaxy magnitude depends on the density contrast $\delta$
in regions where they reside, irrespective of the smoothing scale \cite{Falt2010,moyang04}. Here we adopt the linear relation $\Delta m=0.2 \delta$
which is observationally inferred from the visible band \cite{crotL05}. The actual dependence is a function of the photometric band, and it is
expected to be much weaker in Euclid's H-band as indicated by the weak environmental dependence of the Schechter parameter $M_*$ fitted to NIR
luminosities \cite{mercurio12}. The remaining Schechter parameter, usually dubbed $\alpha$, exhibits a stronger dependence, but since this parameter
fixes the shape at the faint end, it will have only a small impact in deep surveys like Euclid. Nonetheless, it will be possible to remove most of the
contamination caused by this effect using the observed distribution of galaxies in the Euclid survey.

In the weak-field limit, the magnification induced by gravitational lensing is proportional to $1+2\kappa$, where $\kappa$ is the effective convergence
field, i.e. an integral over the (weighted) density contrast along the line of sight \cite{BS01}. For a flat $\Lambda$CDM model and a fixed source
redshift corresponding to a comoving source distance $r$, one finds
\begin{equation}
\kappa(\bm{\theta},r) = \frac{3H_{0}^{2}\Omega_{0}}{2c^{2}}\int_{0}^{r}dr^{\prime}\frac{r^{\prime}(r - r^{\prime})}
{a(r^{\prime})r}\delta(r^{\prime}\bm{\theta},r^{\prime}),
\end{equation}
where the two-dimensional angular vector $\bm{\theta}$ is perpendicular to the line of sight.
Therefore, it is straightforward to model the magnification and its effect in our analysis. The magnification field contains valuable information since
it probes the growth of the angular derivatives of the gravitational potential \cite[e.g.][]{broad95,BS01}. It can also be used to constrain the
gravitational slip which arises in certain modifications of the general theory of relativity \cite[e.g.][]{zhang07,bert11}. Given a theory of gravity,
much of the contribution to the power spectra from gravitational lensing can, in principle, be removed using the underlying large-scale density field
which is inferred from the foreground galaxy distribution. Considering the following analysis, however, we will treat the effects of lensing
magnification as part of the signal. As we will see below, its contribution is not negligible, and can be constrained together with the power spectrum
of the velocity field.

\section{Theoretical angular power spectra}
\label{sec:tpk}
In the following, we will present predictions for the angular power spectra corresponding to the various terms in eq. \eqref{eq:Theta}. Since the observed
galaxies cover a redshift range $z_{1} < z < z_{2}$, we do not consider the angular power spectra of a quantity $f$ defined at a specific redshift, but
instead we use
\begin{equation}
\tilde{f}(\hvn) = \int_{r_{1}}^{r_{2}}f\left\lbrack\hvn r, t(r)\right\rbrack p(r) dr,
\end{equation}
where $r_{1}$ and $r_{2}$ are the comoving distances at the survey's limiting redshifts, $z_1$ and $z_2$, respectively, and $p(r)dr$ is the probability of
observing a galaxy within the interval $[r,r+dr]$. Note that the function $f$ represents any of the terms on the right-hand side of eq. \eqref{eq:Theta}, i.e.
$\sv$, $\sw$, and $\sisw$. Beginning with $\sw$, we write 
\begin{equation}
\tsw_{lm} = \int d \Omega \tsw Y_{lm}(\hvn) = \frac{1}{c^{2}}\int d\Omega Y_{lm}(\hvn) \int_{r_1}^{r_2} \wph(r) \Phi_0(\hvn r) dr ,
\end{equation}
where we have used the linear relation $\Phi(\vr,t)=(D/a)\Phi_0(\vr,t_0)$ and defined $\wph = Dp(r)/a$, with $D(t)$ and $a(t)$ evaluated at $t=t(r)$. 
Expanding $\Phi_0(\vr)$ in Fourier space,
\begin{equation}
\Phi_0(\vr) = \frac{1}{(2\pi)^3}\int d^3 k \Phi_{\vk}{\rm e}^{{\rm i}\vk \cdot{  \vr}},
\end{equation}
and using  
\begin{equation}
\label{eq:ekr}
{\rm e}^{{\rm i}\vk \cdot{\vr}} = 4\pi \sum\limits_{l,m}{\rm i}^lj_l(kr) Y^*_{lm}(\hat {\vn})Y_{lm}(\hat {\vk}),
\end{equation}
where $j_{l}$ is the usual first-kind spherical Bessel function of degree $l$, we get 
\begin{equation}
\tsw_{lm} = \frac{{\rm i}^l}{2\pi^2 c^2}\int_{r_{1}}^{r_{2}} dr \wph \int d^3 k \Phi_{\vk}j_l(kr) Y_{lm}(\hat {\vk}). 
\end{equation}
Therefore, we finally arrive at
\begin{equation}
\label{eq:clphi}
C_{l}^{\Phi} = \langle |\tsw _{lm}|^{2}\rangle = \frac{2}{\pi c^4}\int dk k^2 P_\Phi(k) \left\vert \int_{r_1}^{r_2} d r \wph j_l(kr)\right\vert^{2},
\end{equation}
where we have used $\langle\Phi_{\vk}\Phi_{\vk'}\rangle = (2\pi)^3 \delta^D(\vk-\vk')P_\Phi(k)$.
Similarly, using the linear relation in eq. \eqref{eq:linv}, we obtain
\begin{equation}
C_l^{\rm V} = \frac{2}{\pi c^2}\int dk k^2 P_\Phi(k) \left\vert \int_{r_{1}}^{r_{2}} dr \wv \left(\frac{lj_l}{r}-kj_{l+1}\right)\right\vert^{2},
\end{equation}
where $\wv = 2a\dot{D}p(r)/3\Omega_0 H_{0}^{2}$. As for $\sisw$, the last term appearing in eq. \eqref{eq:Theta}, we will assume that the signal
is mostly caused by the large-scale structure between the observed high redshift galaxy sample and the observer. Therefore, $\sisw$ is
approximately the same for all galaxies along a common line-of-sight such that $\tsisw\approx \sisw$. The angular power spectrum then reads
\begin{equation}
\label{eq:isw}
C_{l}^{\dot{\Phi}} = \langle |\sisw_{lm}|^{2}\rangle = \frac{8}{\pi c^6}\int dk k^2 P_\Phi(k) \left\vert \int_{0}^r dr^{\prime} \wpdot j_l(kr^{\prime})\right\vert^{2},
\end{equation}
where $\wpdot(t)/a = (d/dt)[D/a]$. The integration over $r^{\prime}$ is taken from $r=0$ out to a distance beyond which $\Phi$ becomes nearly
constant with time. For simplicity, we will assume that most of the contribution to the integral comes from the inner edge of the considered
galaxy sample, and adopt $z = 1$ in eq. \eqref{eq:isw} for all galaxies in the survey. For Euclid, this seems to be the case since the mean
redshift is expected to be $z\sim 1$. Note that the same assumptions are used when calculating the angular power spectrum $C^{\rm lens}$.
Although not explicitly given, similar expressions may be obtained for the corresponding cross-correlations of the above contributions.

\subsection{Predictions of the $\Lambda$CDM scenario}
Having derived the relevant expressions above, we are now ready to make predictions for the framework of $\Lambda$CDM. Here we adopt a spatially
flat model with best-fit parameters based on the CMB anisotropies measured by the Wilkinson Microwave Anisotropy Probe (WMAP) \cite{wmap7}. In
this case, the total mass density parameter is $\Omega_m=0.266$, the baryonic density parameter $\Omega_b=0.0449$, the Hubble constant $h=0.71$
in units of  $100\kms {\rm Mpc}^{-1}$, the scalar spectral index $n_{s}=0.963$, and $\sigma_{8} = 0.80$ for the rms of linear density fluctuations
within spheres of $8\hmpc$. We work with a parametric form of the power spectrum taken from ref. \cite{EH98} (see eqs. 29--31 in their paper). For
the calculation of angular power spectra, we use the expressions from section \ref{sec:tpk} together with $p(r)$ corresponding to the redshift
distribution of galaxies appropriate for Euclid which is given by eq. \eqref{eq:nz}.

\begin{figure} 
\centering
\includegraphics[scale=0.7]{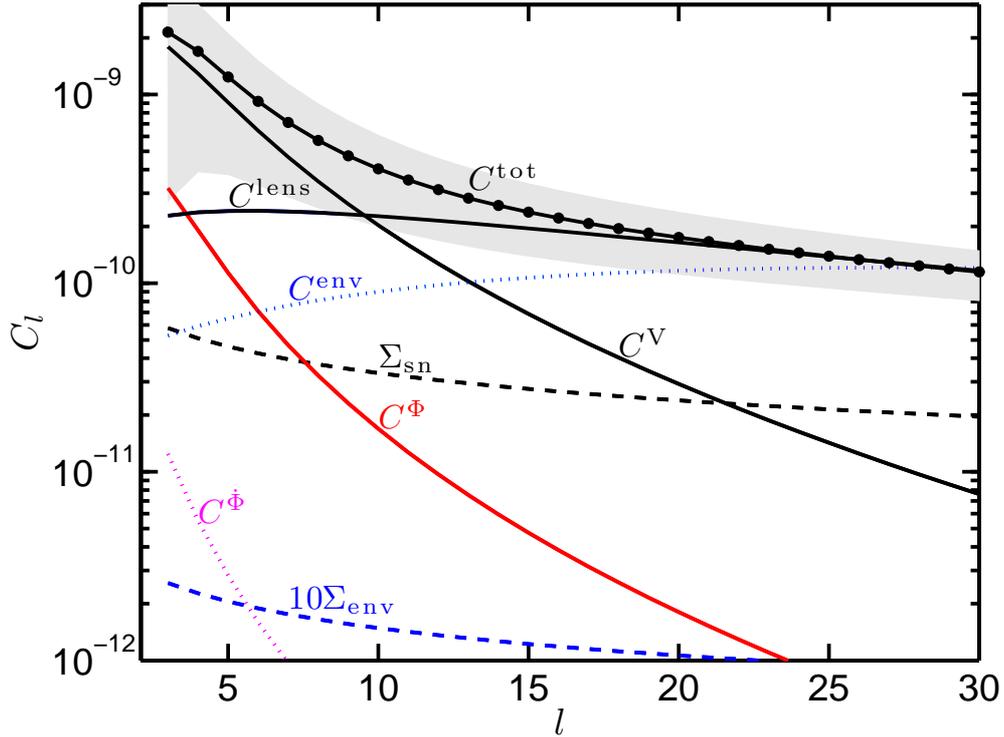}
\caption{Angular power spectra in the $\Lambda$CDM model as explained in the text: The shaded area represents the cosmic variance uncertainty
$\Sigma_{\rm cv}$ on $C^{\rm tot}$, and $\Sigma_{\rm sn}$ is the $1\sigma$ shot-noise error for the Euclid survey. The desired signal is
contaminated by $C^{\rm env}$ which results from environmental dependences in the luminosity function and can be removed from the actual
Euclid data with high precision represented by $\Sigma_{\rm env}$. Only contributions (including cross-correlations) from $\Theta^{\rm V}$
(peculiar velocity), $\Theta^{\Phi}$ (gravitational shift), and $\Theta^{\rm lens}$ (lensing magnification) are included in $C^{\rm tot}$.
At $l\sim 60$ (not shown), $\Sigma_{\rm sn}$ and $C^{\rm tot}$ become comparable.}
\label{fig:two}
\end{figure}

The angular power spectra and associated errors are plotted in figure \ref{fig:two}. The shot-noise error $\Sigma_{\rm sn}$ is
computed using eqs. \eqref{eq:sigi_def} and \eqref{eq:sigf}, which gives $\sigma_{f} = 0.17$ as inferred from the (red) dashed curve in
figure \ref{fig:one}. For completeness, we also show the power spectrum $C^{\rm env}$ which results from environmental dependences in the luminosity function. As explained in section \ref{sec:env}, the
contamination arising from this effect can be removed given the observed distribution of galaxies in the survey. The accuracy to which this can be
achieved for Euclid is represented by $\Sigma_{\rm env}$, the $1\sigma$ error within which $C^{\rm env}$ can be estimated explicitly from the data
(see appendix \ref{sec:shotn} for details). As is clear from $C^{\rm lens}$ in the figure, magnification caused by gravitational lensing introduces
significant angular correlations in $(z_i-\zcos)/(1+z_i)$.

The quantity $C^{\rm tot}$ is the power spectrum of the sum of the three signals $\Theta^{\rm lens}$, $\Theta^{\rm V}$, and $\Theta^\Phi$, 
corresponding to the lensing magnification, the Doppler shift, and the gravitational shift, respectively. Note that the calculation of
$C^{\rm tot}$ does include covariance between the three individual signals. The contribution from $\Theta^{\dot \Phi}$ is negligible as is
indicated by $C^{\dot \Phi}$, and we do not include it in $C^{\rm tot}$.  At $l\ltsim 10$, $C^{\rm V}$ is the dominant
contribution, but the lensing term $C^{\rm lens}$ takes over at $l\gtrsim 10$--$15$, roughly until $l\sim 60$ where it drops below the shot-noise
level. The shaded area represents the $1\sigma$ error of $C^{\rm tot}$ due to cosmic variance. For practical purposes, it is therefore possible
to provide measurements of $C^{\rm V}$ and $C^{\rm lens}$ by an appropriate fitting procedure of the two corresponding curves to the measured
$C^{\rm tot}$. Unfortunately, at low $l$, cosmic variance is so large that accurate constraints on $C^\Phi$ do not seem possible.

\begin{figure} 
\centering
\includegraphics[scale=0.7]{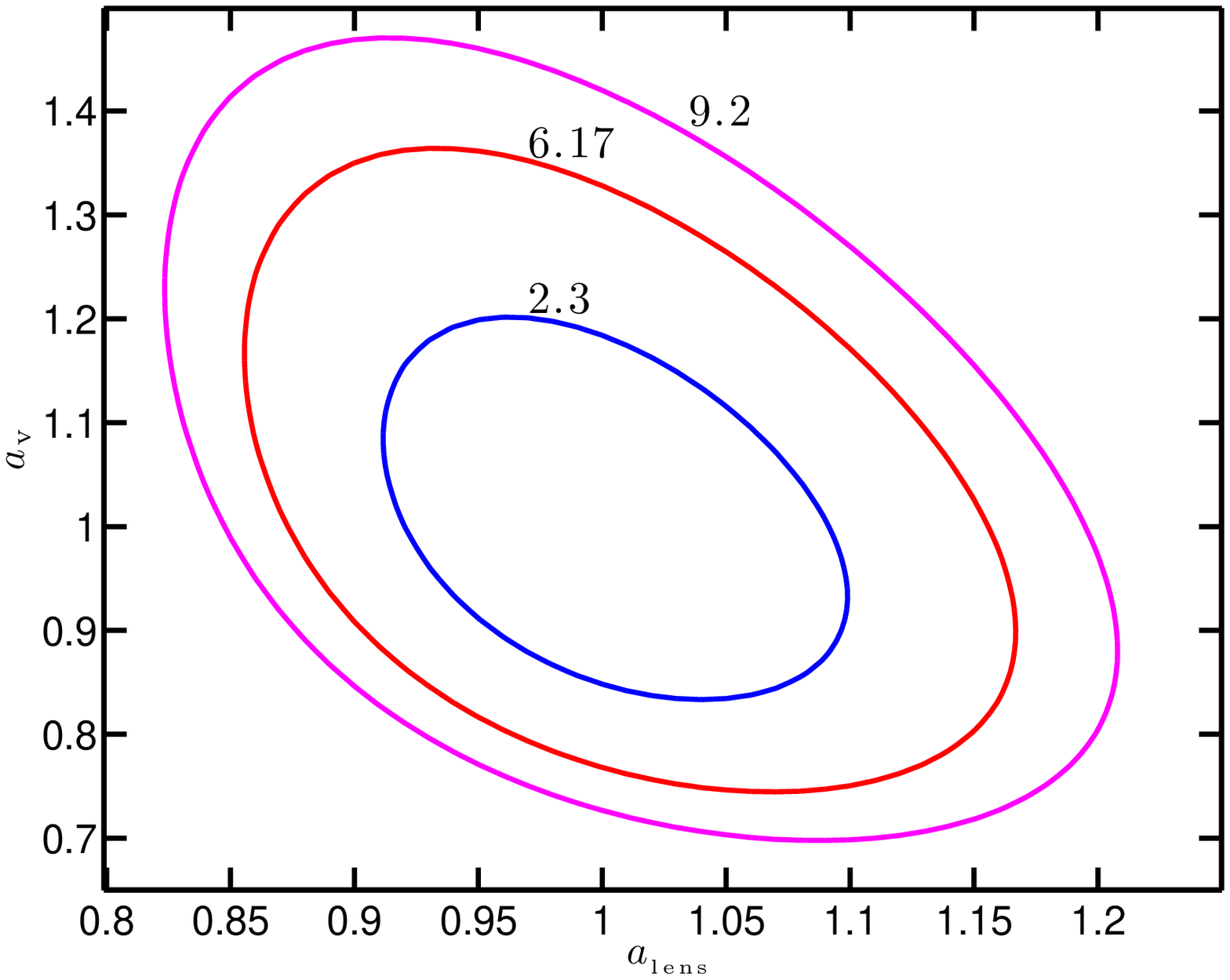}
\caption{Confidence levels for measuring the amplitudes of the velocity and lensing power spectra by fitting
$a_{\rm v}^2 C^{\rm V} + a_{\rm lens}^2 C^{\rm lens}$ to random realizations of $C^{\rm tot}$ which include both shot noise
and cosmic variance: The fitting procedure ignores physical cross-correlations between the used quantities. Contours are shown 
for $\Delta\chi^{2} = 2.3$, $6.17$, and $9.2$, corresponding to confidence levels of 68\%, 95.4\% and 99\%, respectively.}
\label{fig:three}
\end{figure}

\subsection{Signal-to-noise and expected error on model normalization}
Suppose that observations yield an estimate $C^{\rm obs}_{l}$ for the total power spectrum $C^{\rm tot}_{l}$. Let $C_{l}^{\rm H_{0}} = 0$
and $C^{\rm H_{1}}_{l}$ be the expected total power spectra for the null hypothesis with no correlations ($H_{0}$) and of the $\Lambda$CDM
model ($H_{1}$), respectively. The ratio of probabilities for $ H_{1}$ and $H_{0}$ is given by
\begin{equation}
-2\ln\frac{P(H_0)}{P(H_1)} = \sum_l\left[ \frac{\left(C^{\rm obs}_l-C^{\rm H_0}_l\right)^2}{\Sigma_{\rm sn}^2}- \frac{\left(C^{\rm obs}_l-C^{\rm H_1}_l\right)^2}{\Sigma^2}
+2\ln \left(\frac{\Sigma_{\rm sn}}{\Sigma}\right)\right].
\end{equation}
If $H_1$ holds, it follows that $C_l^{\rm H_1}$ is equal to the $C_l^{\rm tot}$ shown in figure \ref{fig:two}, and $C^{\rm obs}_l-C_l^{\rm H_1}$
is a random variable with variance $\Sigma^2$ given by eq. \eqref{eq:SIGMA}. Therefore, the signal-to-noise ratio ($S/N$) for rejecting $H_0$
is
\begin{equation}
\left(\frac{S}{N}\Big\vert_{0}\right)^2 = -2\ln\frac{P(H_0)}{P(H_1)} = \sum_l \left[\frac{(2l+1)\fsk C_l^2} {2\sigma_f^4/\bar n^2}
+\left(1+\frac{\bar nC_l}{\sigma_f^2}\right)^2-2\ln\left(1+\frac{\bar nC_l}{\sigma_f^2}\right) \right] , 
\end{equation}
where  $C_l=C_l^{\rm H_1}=C_l^{\rm tot}$. Substituting the relevant quantities into the above, we obtain $(S/N)\vert_{0}=101$, where the above
sum rapidly converges by $l = 30$. On the other hand, if $H_0$ is true, then the $S/N$ for rejecting $H_1$ is given by
\begin{equation}
\begin{split}
\left(\frac{S}{N}\Big\vert_{1}\right)^2 &= -2\ln\frac{P(H_1)}{P(H_0)}\\
{ } &= \sum_l \left[\frac{(2l+1)\fsk C_l^2} {2(\sigma_f^2/\bar n+C_l)^2}
+\left(1+\frac{\bar nC_l}{\sigma_f^2}\right)^{-2}+2\ln\left(1+\frac{\bar nC_l}{\sigma_f^2}\right) \right ],
\end{split}
\end{equation}
which quickly converges at $l\sim 50$ and yields $(S/N)\vert_{1} = 14.7$. The fact that $(S/N)\vert_{1}\ll (S/N)\vert_{0}$ is a result of
cosmic variance which is zero for the $H_{0}$ hypothesis, but significant in the $\Lambda$CDM scenario, i.e.
$\Sigma_{\rm cv}^{2} = 2C_{l}^{2}/(2l+1)\fsk$. 

In addition to $S/N$ considerations, we can assess how well a measurement of $C^{\rm tot}$ would constrain the normalization of the
$\Lambda$CDM power spectrum in terms of $\sigma_8$. To this end, we write the model's total power spectrum as
$C^{\rm tot,m} = (\sigma_8/0.8)^2C^{\rm tot}$, where $C^{\rm tot}$ as illustrated in figure \ref{fig:two} is obtained for $\sigma_{8}=0.8$. The
expected $1\sigma$ error on $\sigma_{8}$ is then $(-\partial^2 \ln P(H_1)/\partial \sigma_{8}^{2})^{-1}$, where $P(H_1)$ is now expressed as
\begin{equation}
\label{eq:phone}
\ln P(H_{1}) = -\sum_l\frac{\left(C_l^{\rm obs}- C_l^{\rm tot, m}\right)^2  }{2\Sigma_l^2}-\sum_l\ln \Sigma_{l},
\end{equation}
and $\Sigma_l$ is given by eq. \eqref{eq:SIGMA} with $C_l=C_l^{\rm tot,m}$. Using a normal probability distribution of the form given in eq.
\eqref{eq:phone} with fixed $\sigma_{8} = 0.8$, we have generated $1000$ random realizations of $C^{\rm obs}$. For each of these realizations,
we have maximized $P(H_{1})$ given by eq. \eqref{eq:phone} with respect to $\sigma_8$. As a result, we find that the true value $\sigma_{8}=0.8$
is recovered within a relative $1\sigma$ error of less that $4\%$, without any statistically significant bias. 

We have also inspected the possibility of constraining the velocity and lensing signal amplitudes with the help of the two-parameter model
\begin{equation}
C^{\rm tot, m} = a_{\rm V}^2 C^{\rm V} + a_{\rm lens}^2 C^{\rm lens}.
\end{equation}
This model for $C^{\rm tot, m}$ neglects the contribution of $\Theta^{\Phi}$ to $C^{\rm obs}$ as well as any covariance between the remaining
signals, $\Theta^{\rm V}$ and lensing magnification $\Theta^{\rm lens}$. Using the above expression, we have repeated the procedure described
above for constraining $\sigma_8$ with the $1000$ random realizations of $C^{\rm obs}$. The result is presented in figure \ref{fig:three} which
shows contours of $\Delta \chi^2=-2\ln P+2\ln \max(P)$ as a function of $a_{\rm V}$ and $a_{\rm lens}$.
The contours have been computed for one of the random realizations, but the values of the best-fit parameters (giving the lowest $\Delta\chi^2$)
obtained for this particular realization have been shifted to their underlying value, i.e. unity for both. This is reasonable since the average
best-fit parameters of $1000$ realizations are essentially unbiased. As seen from the figure, $a_{\rm lens} $ is constrained with better
accuracy than the velocity amplitude $a_{\rm V}$. This is consistent with figure \ref{fig:one} which shows that $C^{\rm lens}$ dominates the total
signal over a large range of $l$ while $C^{\rm V}$ is significant only for low $l$ where cosmic variance becomes increasingly important. Still,
both parameters are constrained with good accuracy.

\section{Discussion}
\label{sec:cnl} 
In this paper, we have presented a novel method for deriving direct constraints on the peculiar velocity and gravitational potential power
spectra from currently planned galaxy redshift surveys. The large number of galaxies with photometric redshifts in these surveys allows
one to exploit apparent galaxy magnitudes as a proxy for their cosmological redshifts since it beats down the large scatter in the $\zcos(m)$
relation and the uncertainty in the photometric redshifts. The method aims at directly constraining power spectra of the underlying
fluctuation fields independent of the way galaxies trace mass. Other methods for extracting cosmological information from redshift surveys
rely on accurate measurements of the galaxy power spectrum in redshift space (since galaxy distances remain unknown). The power spectrum and
other statistical measures based on the distribution of galaxies have been successful at probing the nature of dark matter and placing 
important constraints on neutrino masses \cite{tegmark04,deputter12,sanchez12}. Having said that, however, they depend on a very accurate
knowledge of the relation between galaxies and the full underlying matter distribution.
The method we have considered here is less precise, but it is completely independent of the galaxy formation process and offers a much more
sensitive assessment of the underlying physical mechanism driving cosmic acceleration and structure formation. This approach is particularly
worthwhile if such constraints on the velocity field and the gravitational potential are contrasted with local constraints obtained from
data at low redshifts. For example, peculiar motions of galaxies within a distance of $\sim 100\hmpc$ can be measured using tight relations
between intrinsic observables of galaxies \cite[e.g.][]{TF77,springsixdf}, and also using astrometric observations of the Gaia space mission
which is currently scheduled for launch in $2013$. These peculiar motions of galaxies have been useful for constraining cosmological
parameters \citep{DN10} as well as the amplitude of the velocity field in the nearby Universe \citep{ND11a,bilicki11}. 

Although we have presented predictions for the Euclid survey, the science proposed in this paper will not have to wait for this space
mission. In fact, several ground-based photometric surveys in the optical and near-infrared bands, which will constitute the backbone of
Euclid's photometry, will provide photometric redshift catalogs that can be used for our purposes well before the launch of the satellite.
On a shorter timescale, the VLT Survey Telescope (VST) will be used to carry out the Kilo Degree Survey (KiDS), one of the ESO public
surveys. It will cover 1,500 deg$^2$  to $u=24$, $g=24.6$, $r=24.4$, and $i=23.1$, and will probably contain $\sim 10^{8}$ galaxies with
measured photometric redshifts.\footnote{\protect\url{http://www.eso.org/public/teles-instr/surveytelescopes/vst.html}} Also, the Dark
Energy Survey (DES) will start its operations soon, and it will cover 5000 deg$^2$ of the Southern sky within 5 years, reaching magnitudes
up to $\sim 24$ in SDSS $griz$ filters, comparable to the limiting magnitudes of Euclid and with a redshift distribution $dN/dz$ similar
to that of Euclid galaxies. DES will measure photometric redshifts of $\sim 3\times 10^8 $ galaxies with $\sigma_{\rm photo}\sim 0.12$ at
$z\sim 1$.\footnote{\protect\url{http://www.darkenergysurvey.org/reports/proposal-standalone.ps}} Furthermore, the first of the four
planned Pan-STARRS telescopes has been operational since May 2010.\footnote{\protect\url{http://pan-starrs.ifa.hawaii.edu/public/home.html}}
The planned 3$\pi$ area of the sky will be considerably shallower, detecting galaxies below a limiting
magnitude of $\sim 24$ in the $griz$ bands. A deeper survey involving the PS1 and PS2 telescopes is currently being planned. The survey
should cover $\sim 7,500 \rm deg^2$ with limiting fluxes $g = 24.7$, $r = 24.3$, $i = 24.1$, and $z = 23.6$. Photometric redshifts will
then be measured for $\sim 4.5 \times 10^8 $ galaxies with similar errors.
Finally, on the long run, the Large Synoptic Survey Telescope (LSST) is expected to start operations in 2020. Its main deep-wide-fast survey
is expected to observe $\sim 20,000$ deg$^2$ in the $ugrizy$ bands. After about 10 years of operation, it will reach much deeper depths
(down to a co-added magnitude $r=27$), detecting about $3 \times 10^9$ galaxies \cite{ivezic08}.

\acknowledgments

We are grateful to Micol Bolzonella for computing all redshift-magnitude relations used in this paper from the zCOSMOS data
and for running simulations to assess the amplitude of zeropoint errors on the measured photometric redshifts.
We also thank Henry McCracken for providing us with the H-band magnitudes of zCOSMOS galaxies. E.B. thanks Gianni Zamorani and
Massimo Meneghetti for useful discussions and suggestions. This work was supported by THE ISRAEL SCIENCE FOUNDATION (grant No.203/09),
the German-Israeli Foundation for Research and Development, the Asher Space Research Institute and the  WINNIPEG  RESEARCH FUND.
E.B. acknowledges the support provided by MIUR PRIN 2008 ``Dark energy and cosmology with large galaxy surveys'' and by Agenzia
Spaziale Italiana (ASI-Uni Bologna-Astronomy Dept. 'Euclid-NIS' I/039/10/0). M.F. is supported in part at the Technion by the
Lady Davis Foundation.

\appendix

\section{Proof of the relations in eq. (\ref{eq:CLC})}
\label{sec:proof}
For any function $f(\hvn)$, we have defined 
\begin{equation}
f_{lm} = \frac{1}{\fsk^{1/2}} \int d\Omega f(\hvn) Y_{lm}(\hvn)\; .
\end{equation}
Using the relations given in eq. (\ref{eq:rules}), we get  
\begin{equation}
\begin{split}
C_{l} & = \frac{1}{(2l+1)\fsk}\sum_{m}\langle\left\vert f_{lm}\right\vert^{2}\rangle\\
& = \frac{1}{(2l+1)\fsk^2}\int d\Omega d\Omega^{\prime}\langle f(\hvn)f(\hvn')\rangle\sum_{m}Y_{lm}(\hvn)Y^*_{lm}(\hvn^{\prime})\\
& = \frac{1}{4\pi\fsk} \int d\Omega d\Omega^{\prime} C(\hvn\cdot \hvn^{\prime})P_l(\hvn\cdot\hvn^{\prime})\\
& = \frac{1}{4\pi\fsk}\int d\Omega^{\prime}  \int  d\Omega C(\hvn\cdot \hvn^{\prime})P_l(\hvn\cdot\hvn^{\prime})\\
& = \int  d\Omega C(\cos\theta)P_l(\cos\theta),
\end{split}
\end{equation}
where for the last step, we have used that the integral over $d\Omega$ is independent of $\hvn'$, assuming that the coherence angular length
is smaller than the spatial extent of the survey. Similarly, the first relation in eq. (\ref{eq:CLC}) follows from substituting it into the
second and exploiting the (near) orthogonality of $P_{l}(\cos\theta)$.

\section{Estimating angular power spectra for discrete data}
\label{sec:shotn}
Just as in the main body of the paper, we will take the symbol $\langle (\cdot )\rangle_{_{m}} \equiv (1/(2l+1)\fsk)\sum_m (\cdot)$ to denote averaging over
all indices $m$ corresponding to the degree $l$ of the spherical harmonic decomposition. Again, the symbol $\langle\cdot\rangle$ without any subscript will
refer to the previously introduced ensemble average, and we shall interchange between $\langle\cdot\rangle_{_{m}}$ and $\langle\cdot\rangle$ whenever appropriate. For
brevity of notation, we also use the definition $Y_{m}^i \equiv Y_{lm}(\hvn_i)$, thus omitting the subscript $l$ since all calculations will refer
to a given degree $l$ of the spherical harmonics. Starting from the definition of $f_{lm}$ in eq. \eqref{eq:fdef}, we obtain
\begin{equation}
\label{eq:SNauto}
\begin{split}
\langle\left| f_{lm}\right|^{2}\rangle &= \frac{1}{{\bar n}^2 \fsk} \sum_{i,j}\langle f_i f_j\rangle Y^i_{m} Y^{*j}_{m}\\
&= \frac{1}{{\bar n}^2 \fsk} \sum_{i}\langle f_i^2\rangle\left\vert Y^i_{m}\right\vert^2 + \frac{1}{{\bar n}^2 \fsk} \sum_{i\ne j}\langle f_i f_j\rangle Y^i_{m} Y^{*j}_{m}\\
&= \frac{\sigma_{f}^{2}}{\bar n} + C_{l},
\end{split}
\end{equation}
where the value of $\sigma_{f}$ is inferred from the $\zcos(m)$ relation, i.e. $\sigma_{f}^{2} = \sum_{i}\sigma_{i}^{2}/N$. As our estimate for
$C_{l}$, we therefore take  
\begin{equation}
\label{eq:Cest}
C_{l} = \langle\left| f_{lm}\right |^{2}\rangle_{_{m}} - \frac{\sigma_f^2}{\bar n}.
\end{equation}
Another way of arriving at this result is to partition the observed sky into infinitesimally small cells of angular size $\delta\Omega$, with each
cell containing at most one galaxy \citep{Peeb80}. In this case, we write 
\begin{equation}
f_{lm} = \frac{1}{{\bar n} \fsk^{1/2}}\sum_\alpha f_\alpha n_\alpha Y^\alpha_{m},
\end{equation}
where $n_\alpha$, the number of galaxies in cell $\alpha$, is either 0 or 1. Using that $\langle n_\alpha\rangle =\bar{n}\delta\Omega$ in this representation,
we obtain the expression 
\begin{equation}
\begin{split}
\langle\left |f_{lm}\right |^{2}\rangle &= \frac{1}{{\bar n}^2\fsk }\sum_{\alpha,\beta}\langle f_{\alpha}f_{\beta}\rangle\langle n_{\alpha}n_{\beta}\rangle Y^\alpha_{m}Y^{*\beta}_{m} \\
&= \frac{1}{{\bar{n}}^2\fsk }\sum_{\alpha}\langle f_{\alpha}^{2}\rangle\langle n_{\alpha}^{2}\rangle\left\vert Y^{\alpha}_{m}\right\vert^{2} + C_{l}\\
&= \frac{1}{{\bar{n}}^2\fsk }\sum_{\alpha}\langle f_{\alpha}^{2}\rangle\bar{n}\delta\Omega \left\vert Y^{\alpha}_{m}\right\vert^{2} + C_{l}\\
&= \frac{\sigma_{f}^{2}}{\bar{n}} + C_{l},
\end{split}
\end{equation}
where we have used the relation $n_{\alpha}^{2} = n_{\alpha}$. 
We find this particular approach a little more intuitive to compute the expected error of the estimate in eq. \eqref{eq:Cest}. Let $f^r$ and $n^r$
be possible random realizations of the data in another universe. The variance of the error in $C_{l}$ may then be written as an ensemble average
over these realizations, i.e.
\begin{equation}
\label{eq:erest}
\begin{split}
\Sigma^{2} &= \langle (C_{l}^{r} - C_{l})^{2}\rangle\\ 
&= \left\langle\left (\frac{1}{{\bar{n}}^{2}\fsk}\sum_{\alpha\ne\beta}f_{\alpha}^{r}f_{\beta}^{r}n_{\alpha}^{r}n_{\beta}^{r} 
\langle Y_{m}^{\alpha}Y_{m}^{*\beta}\rangle_{_m} - C_{l}\right )^{2}\right\rangle\\
&= \frac{1}{{\bar{n}}^{4}\fsk^{2}}\sum_{\alpha\ne\beta,\alpha^{\prime}\ne\beta^{\prime}}\langle f_{\alpha}^{r}n_{\alpha}^{r}f_{\beta}^{r}n_{\beta}^{r}f_{\alpha^{\prime}}^{r}n_{\alpha^{\prime}}^{r}
f_{\beta^{\prime}}^{r}n_{\beta^{\prime}}^{r}\rangle
\langle Y_{m}^{\alpha}Y_{m}^{*\beta}\rangle_{_m}\langle Y_{m^{\prime}}^{\alpha^{\prime}}Y_{m^{\prime}}^{*\beta^{\prime}}\rangle_{_{m^{\prime}}} - C_{l}^{2}.
\end{split}
\end{equation}
Note the remarkable fact that due to the uncorrelated nature of the errors and because $\alpha\ne\beta$ as well as $\alpha'\ne \beta'$, only
second-order moments of $n$ and $f$ contribute to $\Sigma^2$. This is important since it implies that the non-Gaussian nature of the scatter
in the $\zcos(m)$ relation does not affect the variance of errors in the estimated $C_{l}$. Assuming that $fn$ is a Gaussian random field, we
further have 
\begin{equation}
\label{eq:aux}
\begin{split}
\langle f_{\alpha}^{r}n_{\alpha}^{r}f_{\beta}^{r}n_{\beta}^{r}f_{\alpha^{\prime}}^{r}n_{\alpha^{\prime}}^{r}
f_{\beta^{\prime}}^{r}n_{\beta^{\prime}}^{r}\rangle = &\langle f_{\alpha}^{r}f_{\beta}^{r}n_{\alpha}^{r}n_{\beta}^{r}\rangle\langle f_{\alpha^{\prime}}^{r}f_{\beta^{\prime}}^{r}n_{\alpha^{\prime}}^{r}n_{\beta^{\prime}}^{r}\rangle\\
&+ \langle f_{\alpha}^{r}f_{\alpha^{\prime}}^{r}n_{\alpha}^{r}n_{\alpha^{\prime}}^{r}\rangle\langle f_{\beta}^{r}f_{\beta^{\prime}}^{r}n_{\beta}^{r}n_{\beta^{\prime}}^{r}\rangle\\
&+ \langle f_{\alpha}^{r}f_{\beta^{\prime}}^{r}n_{\alpha}^{r}n_{\beta^{\prime}}^{r}\rangle\langle f_{\alpha^{\prime}}^{r}f_{\beta}^{r}n_{\alpha^{\prime}}^{r}n_{\beta}^{r}\rangle .
\end{split}
\end{equation}
Considering the above, averages with no equal indices will contribute terms proportional to $C_{l}^{2}$ in the sum of eq. \eqref{eq:erest}.
Neglecting the contribution of shot noise, this would yield the usual cosmic variance expression $2C_{l}^{2}/(2l+1)\fsk$. For the case
$\alpha=\alpha^{\prime}$ and $\beta=\beta^{\prime}$, the second term on the right-hand side of eq. \eqref{eq:aux} turns into
$\sigma_{f}^{4}(\bar n \delta \Omega)^2$, where we have again used that $\langle n_{\alpha}^{2}\rangle = \langle n_{\alpha}\rangle = \bar{n}\delta\Omega$. The other
combination which makes the same contribution is $\alpha=\beta^{\prime}$ and $\alpha^{\prime}=\beta$. A bit of algebra shows that the
contribution to the variance in eq. \eqref{eq:erest} of these two combinations is $2\sigma_{f}^{4}/(2l+1)\bar{n}^{2}\fsk$. For
$\alpha=\alpha^{\prime}$ and $\beta\ne\beta^{\prime}$, the second term on the right-hand side of eq. \eqref{eq:aux} simplifies to
$\sigma_{f}^{2} \langle f^{r}_{\beta}f^{r}_{\beta\prime}\rangle (\bar{n}\delta\Omega)^{3}$, and the total result of similar permutations with only
two equal indices reads $4\sigma_{f}^{2}C_{l}/(2l+1)\bar{n}\fsk$. Finally, summing up all relevant terms leads to 
\begin{equation}
\Sigma^{2} = \frac{2}{(2l+1)\fsk}\left(\frac{\sigma_{f}^{2}}{\bar{n}} + C_{l}\right)^{2}.
\end{equation}
Using the weights defined in eq. \eqref{eq:w}, it is easy to say that the same results are valid with $\sigma_{f}$ given by eq.
\eqref{eq:sigf}.

In the main body of the paper, we have assumed that environmental dependences of the galaxy luminosity function arise due to variations in
the large-scale density contrast. The signal contamination associated with this effect has an angular power spectrum which is proportional
to that of the density contrast. In the following, we give an estimate for this power spectrum and the $1\sigma$ error within which it can
be constrained from the observed galaxy distribution. To begin with, we consider \cite{Peeb80}
\begin{equation}
\delta_{lm} = \frac{1}{\bar{n}\fsk^{1/2}}\sum_{\alpha}n_{\alpha}Y^{\alpha}_{m}.
\end{equation}
For the $f_{lm}$'s, we have assumed that $n_{\alpha}$ is a discrete sampling of a uniform distribution. That was consistent with linear
theory since we have assumed that $f$ is a function of the density contrast. However, now we are interested in the autocorrelation function
of $n_{\alpha}$, and thus the treatment of $\delta_{lm}$ is a little different. To obtain an estimate for the angular power spectrum of the
density contrast, we write
\begin{equation}
\begin{split}
\langle\vert\delta_{lm}\vert^{2}\rangle &= \frac{1}{\bar{n}^{2}\fsk}\sum_{\alpha,\beta}\langle n_{\alpha}n_{\beta}\rangle Y_{m}^{\alpha}Y^{\beta}_{m}\\
{ } &= \frac{1}{\bar{n}} + C^{\delta\delta}_{l}.
\end{split}
\end{equation}
Similar as before, we hence take $C^{\delta\delta}_{l} = \langle\vert\delta_{lm}\vert^{2}\rangle_{m} - 1/\bar{n}$ as our estimate of the angular power
spectrum of $\delta$. The statistical error in this estimate can be derived using the same approach as above, and it is given by
\begin{equation}
\label{eq:errdd}
\Sigma_{\delta\delta}^{2} = \frac{2}{(2l+1)\fsk}\left(\frac{1}{\bar{n}} + C_{l}^{\delta \delta}\right)^{2}.
\end{equation}
One of the main interests in this paper is to remove the contamination $C_{l}^{\rm env}$ to the signal $C_{l}^{\rm tot}$ using the actually
observed distribution of galaxies. Note that in this case, cosmic variance is irrelevant to the analysis since we are referring to a given
data set rather than the expected deviation of the estimated power spectrum from some underlying theoretical model. Therefore, we may
simply neglect the contribution of cosmic variance in eq. \eqref{eq:errdd}, and the variance of the relevant error reduces to the expression
$2/(2l+1)\bar{n}\fsk$.

\bibliography{gISW}
\end{document}